\documentclass[prb,twocolumn,preprintnumbers,amsmath,amssymb,superscriptaddress,showpacs]{revtex4}
\usepackage{graphicx}
\usepackage{epsfig}
\usepackage{dcolumn} 
\usepackage{bm} 

\begin{document}

\title{Magnetodielectric effect in the $S$ = 1/2 quasi-two dimensional antiferromagnet
K$_2$V$_3$O$_8$}

\author{{R. C. Rai}}
\affiliation{Department of Chemistry, University of Tennessee,
Knoxville, TN 37996}
\author{{J. Cao}}
\affiliation{Department of Chemistry, University of Tennessee,
Knoxville, TN 37996}
\author{J. L. Musfeldt}
\affiliation{Department of Chemistry, University of Tennessee,
Knoxville, TN 37996}
\author{{D. J. Singh}}
\affiliation{Oak Ridge National Laboratory, P. O. Box 2008, Oak
Ridge, Tennessee 37831}
\author{{X. Wei}}
\affiliation{National High Magnetic Field Laboratory, Florida State
University, Tallahassee, Florida 32310}
\author{{R. Jin}}
\affiliation{Oak Ridge National Laboratory, P. O. Box 2008, Oak
Ridge, Tennessee 37831}
\author{{Z. X. Zhou}}
\affiliation{Oak Ridge National Laboratory, P. O. Box 2008, Oak
Ridge, Tennessee 37831}
\author{{B. C. Sales}}
\affiliation{Oak Ridge National Laboratory, P. O. Box 2008, Oak
Ridge, Tennessee 37831}
\author{{D. Mandrus}}
\affiliation{Oak Ridge National Laboratory, P. O. Box 2008, Oak
Ridge, Tennessee 37831}

\begin{abstract}

We report the optical and magneto-optical properties of
K$_2$V$_3$O$_8$, an $S$=1/2 quasi-two-dimensional Heisenberg
antiferromagnet. Local spin density approximation electronic
structure calculations are used to assign the observed excitations
and analyze the field dependent features. Two large magneto-optical
effects, centered at $\sim$1.19 and 2.5 eV, are attributed to
field-induced changes in the $V^{4+}$ $d$ $\rightarrow$ $d$ on-site
excitations due to modification of the local crystal field
environment of the VO$_5$ square pyramids with applied magnetic
field. Taken together, the evidence for a soft lattice, the presence
of vibrational fine structure on the sharp 1.19 eV magneto-optical
feature, and the fact that these optical excitations are due to
transitions from a nearly pure spin polarized V $d$ state to
hybridized states involving both V and O, suggest that the
magneto-dielectric effect in K$_2$V$_3$O$_8$ is driven by strong
lattice coupling.

\end{abstract}




\maketitle \clearpage

\section{INTRODUCTION}

Thermochromic, electrochromic, piezochromic, and photochromic
materials have attracted a great deal of attention in recent
years\cite{Casado2004,DeLongchamp2004,Khan1991,Rodriguez2000,Hayashi1995,
Wild1973,Koidl1976,Faughnan1971} due to the compelling underlying
physics as well as possible device applications. As a simple example
of physical tuning, HgI$_2$ changes color (from red to yellow) with
temperature due to an $\alpha$- to $\beta$-phase structural
transition.\cite{Hill2005} At least one Cu$_2$(radical-ligand)$_2$
complex shows thermochromism, changing from brown-black to green
with decreasing temperature.\cite{Lahti} Electrochromic effects have
been observed in a number of conjugated polymers\cite{Argun2004} and
density wave materials,\cite{Brill94} effects attributed  to
doping/dedoping and 
polarization modifications. An applied electric field can also
provide reversible modulation of the magnetic moment and coercive
field, as demonstrated in Co-doped anatase TiO$_2$-based
devices,\cite{Zhao2005} and magnetic phase control, as illustrated
by recent work on HoMnO$_3$.\cite{Lotter2004} Hydrostatic pressure
and shear stress can change bond distances and angles, modifying the
crystal field environment around a transition metal site. For
instance, with increasing pressure, Pt(diphenylglyoximato)$_2$ turns
from red-brown to green above 0.4 GPa; upon rotation of the anvil,
the green film turns yellow.\cite{Inoku2005,Shiro2003} In the area
of photochromism, a  photoinduced insulator-to-metal transition has
been observed in the low-bandwidth manganite
Pr$_{0.7}$Ca$_{0.3}$MnO$_3$,\cite{Miyano1997} and photoinduced
reflectance changes combined with charge order domain switching have
been discovered in Bi$_{0.3}$Ca$_{0.7}$MnO$_3$.\cite{Smoly2001}

\begin{figure}[b]
\includegraphics[width=2.8in]{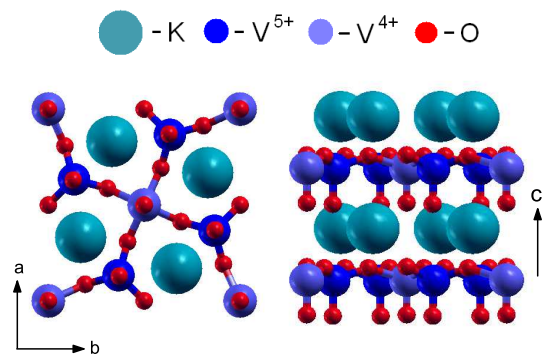}
\vskip -0.2cm \caption{(Color online) Two views of the
K$_2$V$_3$O$_8$ crystal structure. View along c-axis (Left) and view
along b-axis (Right).}\label{struct}
\end{figure}

An applied magnetic field can also induce dielectric property
changes. Low frequency magneto-dielectric effects have been reported
in garnet, HoMnO$_3$, DyMnO$_3$, and DyMn$_2$O$_5$ due to
cross-coupling
effects.\cite{cheong04,Cruz2005,Lorenz2004,TGoto2004,NHur2004}
Ferromagnetic SeCuO$_3$ displays a giant magneto-dielectric and
magneto-capacitance response due to coupling of magnetic
fluctuations to optical phonons.\cite{Lawes2003} Strong coupling
between magnetism and dielectric property is also observed in
La$_2$NiMnO$_6$. \cite{Nyris2005} Higher frequency dielectric
changes are also of interest. With a handful of recent reports on
magnetic field-induced color changes in low-dimensional materials
such as Li$_{0.9}$Mo$_6$O$_{17}$, (CPA)$_2$CuBr$_4$, and
(La$_{0.4}$Pr$_{0.6}$)$_{1.2}$Sr$_{1.8}$Mn$_2$O$_7$,\cite{Choia2004,Choib2004,Woodward}
it is clearly useful to elucidate the operative mechanisms, increase
the size of the magneto-dielectric effect, and assert
molecular-level control over the important coupling processes.
K$_2$V$_3$O$_8$ offers an opportunity to explore how structural and
electronic degrees of freedom can be coupled in layered,
inhomogeneously mixed-valent systems. Photo-induced magnetism,
observed for instance in Prussian blue and related
derivatives,\cite{Sato1996,Kawa1999,Verd1999,Moore2003,Peja2000,Berli2004,Miller2000}
Mn(TCNE)$_x$$\cdot$y(CH$_2$Cl$_2$) and V(TCNE)$_2$ molecular
magnets,\cite{Peja2002,Girtu2000} and dillute magnetic alloy
semiconductors such as (Ga, Mn)As thin films\cite{Oiwa02}
complements on-going work on magneto-dielectric effects. It is also
of interest to manipulate ferromagnetism with light in the junction
geometry\cite{zutic2004} and to combine properties as in the
multiferroic systems.\cite{Kimura2003,vanaken04}

 Figure \ref{struct} displays the tetragonal crystal structure of
K$_2$V$_3$O$_8$.\cite{CARPY1974} It consists of layers of
corner-sharing VO$_5$ square pyramids with magnetic $V^{4+}$ ($S$ =
1/2 ) ions and VO$_4$ tetrahedra with nonmagnetic $V^{5+}$ ($S$ = 0)
ions, separated by interstitial $K^+$ ions. K$_2$V$_3$O$_8$ displays
a structural phase transition at $\sim$110 K, driven by a distortion
of the apical oxygen of the VO$_5$ square pyramids and a distortion
of the unit cell along the $b$
direction.\cite{Withers2004,Choi2001,Sales2002} A weaker basal plane
distortion was also observed near 60 K.\cite{Choi2001} From the
magnetic properties point of view, K$_2$V$_3$O$_8$ is an $S$ = 1/2
two-dimensional Heisenberg antiferromagnet, with exchange constant
$J$ = 12.6 K. It undergoes N\'{e}el ordering at $T_N$ = 4 K,
exhibits unusual field-induced spin reorientations (spin-flop for
$H$$\parallel$$c$ and spin-rotation for $H$$\perp$$c$) due to the
interplay between Dzyaloshinskii-Moriya interactions and easy axis
anisotropy,\cite{Lumsden2001} and displays dramatic field-induced
enhancement of thermal conductivity below 6 K.\cite{Sales2002} The
broad peak in  the susceptibility \cite{Liu1995,Lumsden2001} near 15
K is considered to be a classic signature of two-dimensional
short-range spin correlations in the
$ab$-plane.\cite{Hardy03,Chung2004} Recent theoretical work suggests
a scenario where the Dzyaloshinskii-Moriya interaction suppresses
quantum fluctuations and yields an out-of-plane spin canting angle
of the form ${\rm cos}({\theta}) = H/H_S$, where $H_S = 4S(1 +
{\sqrt{1 + D^2}})$ is the saturation field.\cite{Chernyshev05} For
applied field in the $ab$ plane, one of the magnon branches is
predicted to have an unusual dependence on the Dzyaloshinskii-Moriya
coupling that changes with the size of the field, giving rise to a
possible non-analytic behavior of the spin gap.\cite{Chernyshev05}

In order to investigate the interplay between spin, lattice, and
charge degrees of freedom in low-dimensional, inhomogeneously
mixed-valent vanadates, we measured the optical and magneto-optical
properties of K$_2$V$_3$O$_8$. We observed two prominent
magneto-dielectric effects at base temperature that derive from the
field-induced changes in the local crystal field environment around
the VO$_5$ square pyramid. Combined with evidence for a soft
lattice, the presence of vibrational fine structure on the 1.19 eV
magneto-optical feature, and the fact that these optical excitations
are due to transitions from a nearly pure spin polarized V $d$ state
to hybridized states involving both V and O suggest that the
magneto-dielectric effects in K$_2$V$_3$O$_8$ may be driven by
strong lattice coupling.

\section{Methods}

\subsection{Crystal Growth and Magnetic Characterization}

 Single crystals of K$_2$V$_3$O$_8$ were grown by cooling appropriate
amounts of VO$_2$ in a molten KVO$_3$ flux in a platinum
crucible.\cite{synthesis} Typical crystal dimensions are
$\approx$5$\times$5$\times$1 mm$^3$. Smooth sample surfaces were
prepared by cleaving the crystals parallel to the $ab$-plane and
cleaning with warm water.

High field magnetization studies were carried out using a vibrating
sample magnetometry technique for $H$$\parallel$$c$ and
$H$$\parallel$$ab$. The experiments were performed between 1.6 and
20 K using a 33 T resistive magnet at the National High Magnetic
Field Laboratory (NHMFL) in Tallahassee, FL.

\subsection{Spectroscopic Investigations}

Near normal $ab$-plane  reflectance of K$_2$V$_3$O$_8$ was measured
over a wide energy range (3.7 meV - 6.5 eV) using several different
spectrometers including a Bruker 113 V Fourier transform infrared
spectrometer, a Bruker Equinox 55 Fourier transform infrared
spectrometer equipped with an infrared microscope, and a Perkin
Elmer Lambda 900 grating spectrometer, as described previously.
\cite{zhu2002} The spectral resolution was 2 cm$^{-1}$ in the far
and middle-infrared and 2 nm in the near-infrared, visible, and
near-ultraviolet. Optical conductivity was calculated by a
Kramers-Kronig analysis of the measured
reflectance.\cite{wooten1972} An open flow cryostat and temperature
controller provided temperature control.

The magneto-optical properties of K$_2$V$_3$O$_8$ were investigated
between 0.8 and 3.5 eV using a grating spectrometer equipped with
InGaAs and CCD detectors and a 33 T resistive magnet at the NHMFL.
150 and 600 lines/mm gratings were used, as appropriate. Experiments
were performed between 2.5 and 30 K for $H$$\parallel$$c$. The
field-induced changes in the measured reflectance were studied by
taking the ratio of reflectance at each field and reflectance at
zero field, i.e., [R($H$)/R($H$ = 0 T)]. This normalized response is
a sensitive way to view the field-induced optical
changes.\cite{sushkov2002} To obtain the 30 T optical conductivity,
we renormalized the zero-field absolute reflectance with the
high-field reflectance ratios, and recalculated $\sigma$$_1$ using
Kramers-Kronig techniques.\cite{wooten1972}

\subsection{Electronic Structure Calculations}
 First principles, local spin density approximation (LSDA)
calculations were done in order to better understand the electronic,
magnetic and optical properties of K$_2$V$_3$O$_8$. These were
performed using the general potential linearized augmented planewave
(LAPW) method and with the augmented planewave plus local orbital
modification using well converged basis sets.
\cite{singh-book,sjost,code-note,wien} Local orbital extensions were
employed to include the high lying semi-core states as well as to
relax any linearization errors. \cite{singh-lo} The calculations
included no shape approximations to either the potential or charge
density. Relativity was included for the valence states within a
scalar relativistic approximation, while full relativity was
included for the core states, within an atomic-like approximation.

\section{RESULTS AND DISCUSSION}

\subsection{Understanding the Electronic Structure of K$_2$V$_3$O$_8$}

\begin{figure}[t]
\includegraphics[width=3.0in]{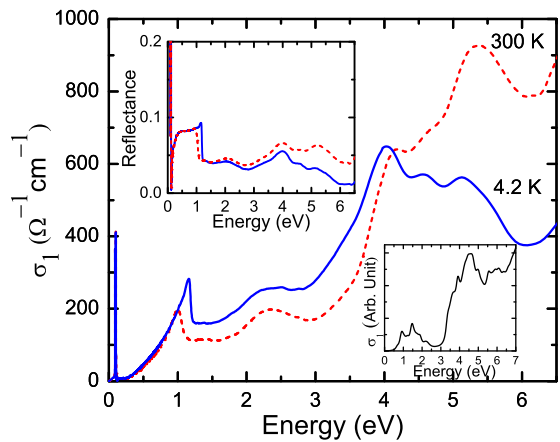}
\vskip -0.2cm \caption{(Color online) The $ab$-plane optical
conductivity spectra of K$_2$V$_3$O$_8$, extracted from the measured
reflectance (upper inset)  by a Kramers-Kronig analysis. The lower
inset shows the calculated optical conductivity, obtained from an
analysis of the electronic structure, with 0.1 eV Lorenztian
broadening. }\label{Conductivity}
\end{figure}

 Figure \ref{Conductivity} displays the $ab$-plane optical
conductivity of K$_2$V$_3$O$_8$. The spectra show strong electronic
and vibrational excitations, characteristic of a semiconductor with
an 0.5 eV optical gap. Based on our electronic structure
calculations (discussed in detail below) and comparison with
chemically similar model
compounds,\cite{Shin1990,Kobayashi1998,Presura2000,Morikawa1995,Tsvetkov2004}
we assign the features centered at $\sim$1 and 2.3 eV to nearly
spin-polarized $V^{4+}$ $d$ $\rightarrow$ $d$ on-site excitations in
the majority spin channel. The 1 eV feature blueshifts with
decreasing temperature and is sensitive to the 110 K structural
phase transition.\cite{Choi2001} We also assign the $\sim$4.0, 4.6,
and 5.2 eV features in the low temperature optical conductivity
spectrum as O 2$p$ $\rightarrow$ V 3$d$ charge transfer excitations.
These features display strong temperature dependence. Partial sum
rule calculations (not shown) indicate that oscillator strength is
conserved up to 5.5 eV, above which the 300 K response is slightly
larger.

\begin{figure}[b!]
\centerline{
\epsfig{file=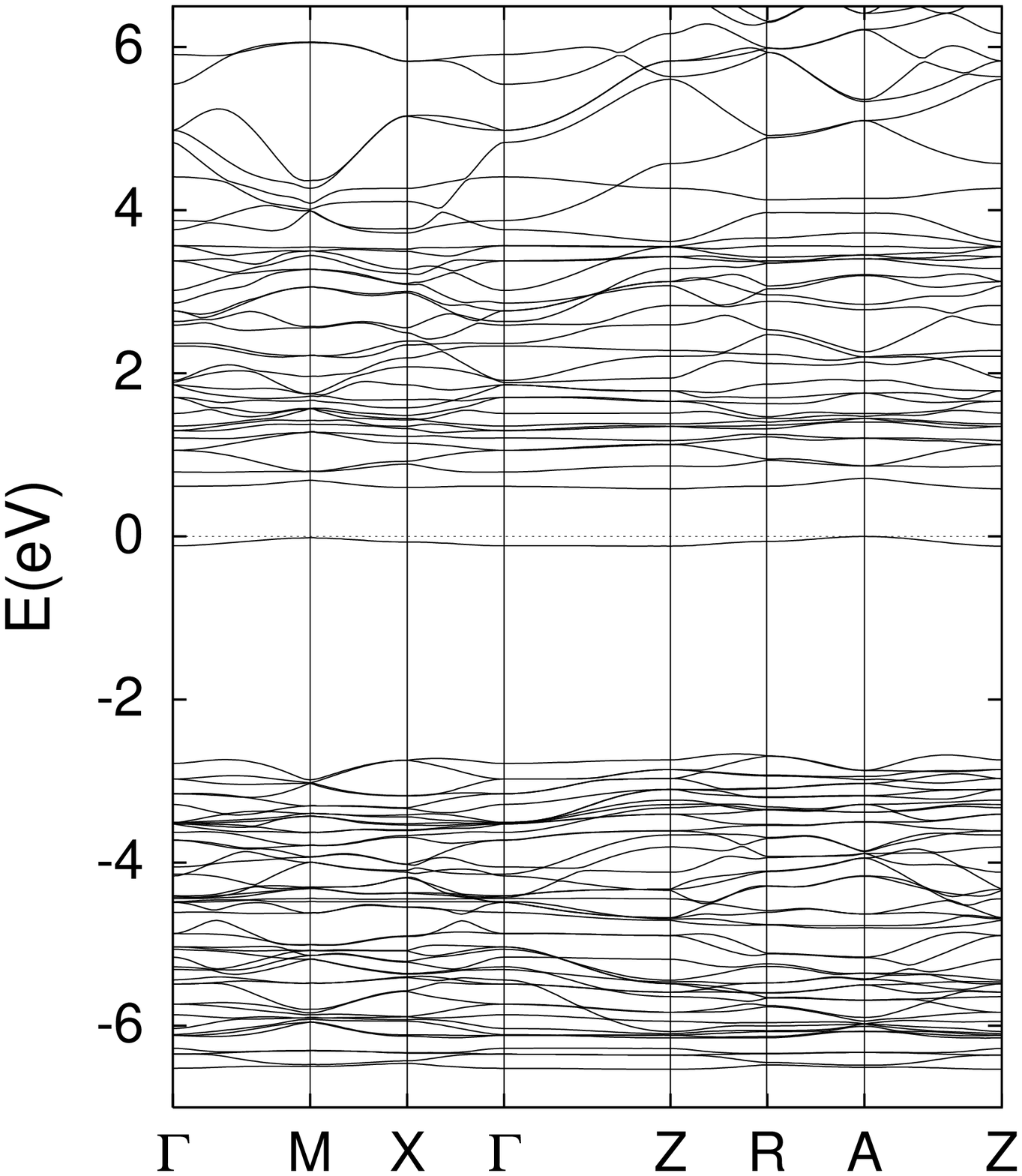,width=0.45\linewidth,angle=0,clip=}
\epsfig{file=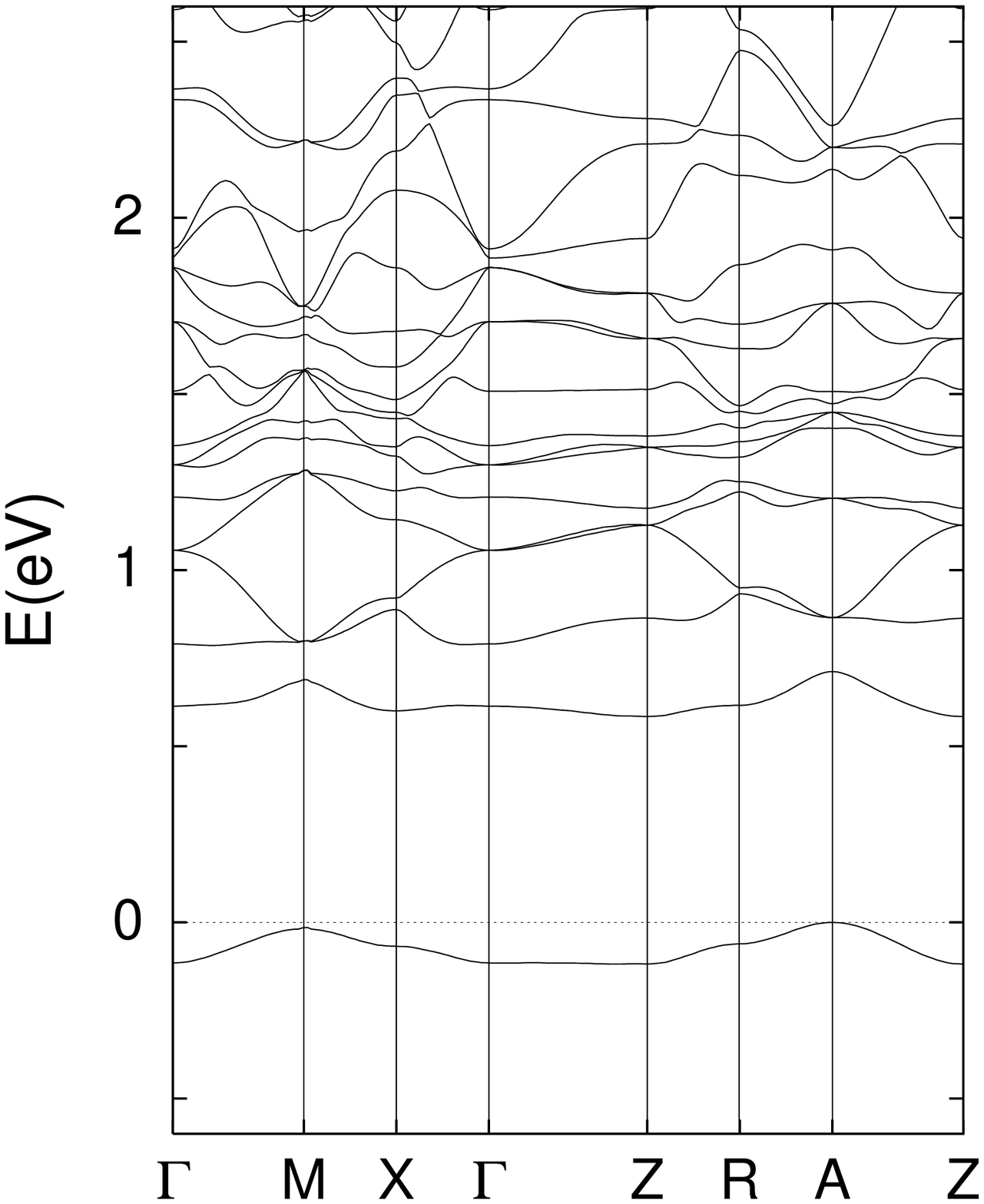,width=0.45\linewidth,angle=0,clip=}}
\vspace{0.2 cm} \caption{LSDA band structure (left) and blow up
(right) for antiferromagnetic K$_2$V$_3$O$_8$. The energy zero is
the highest occupied state.} \label{bands}
\end{figure}

We assigned the observed excitations in the optical conductivity
(Fig. \ref{Conductivity}) based upon our electronic structure
investigations.  The experimental non-centrosymmetric tetragonal
crystal structure (spacegroup $P4bm$, with two formula units per
cell, Fig. \ref{struct}) measured at 120 K was used in these
calculations. This crystal structure has two different V sites, V1
(one per formula unit) and V2 (two per formula unit). The V1 atoms
are closely coordinated by apical O atoms (denoted O1) at a distance
of 2.94 Bohr, forming vanadyl ions. Neutron scattering experiments
show V spins associated with the V1 sites, which at low temperature
order antiferromagnetically in a simple alternating nearest neighbor
pattern within the 2D V-O sheets. \cite{Lumsden2001} This is the
same ground state that is found in the LSDA. Spin polarized
calculations yield local spin moments of 1 $\mu_B$ associated with
the V1 sites  (the insulating ferromagnetic configuration has a spin
moment of exactly 1 $\mu_B$/V1 site; the antiferromagnetic ground
state has the same moment inside the V1 LAPW spheres as the
ferromagnetic case to within 0.002 $\mu_B$). The V2 sites are found
to be in a $d^0$ configuration with no moment as expected. The
energy for the antiferromagnetic configuration is 0.22 eV per
formula unit below that of a constrained non-spin-polarized
calculation, reflecting the Hund's coupling on the vanadyl V1-O1
ions. The ferromagnetic ordered state is 0.055 eV per formula unit
higher than the lowest energy antiferromagnetic state. Both
magnetically ordered states display insulating gaps. Thus strong
local moment character is found. The magnetic exchange energy
corresponding to the difference in energy of the antiferromagnetic
and ferromagnetic states may be an overestimate due to an
overestimation of hopping integrals in the LSDA. However, even
taking this into account, it is very large considering the low
experimental $T_N$. Assuming that the lattice distortion associated
with the 110 or 60 K transitions does not substantially decrease the
exchange coupling, this indicates a substantial reduction of the
ordering temperature due to the strong two dimensionality of the
compound. Thus, especially in-plane, though fluctuating, the V1
moments should be locally antiferromagnetic at temperatures
considerably above $T_N$. This is consistent with the experimental
observation of a susceptibility maximum at $\sim$15 K.
\cite{Liu1995,Lumsden2001}

\begin{figure}[ht]
\centerline{
\epsfig{file=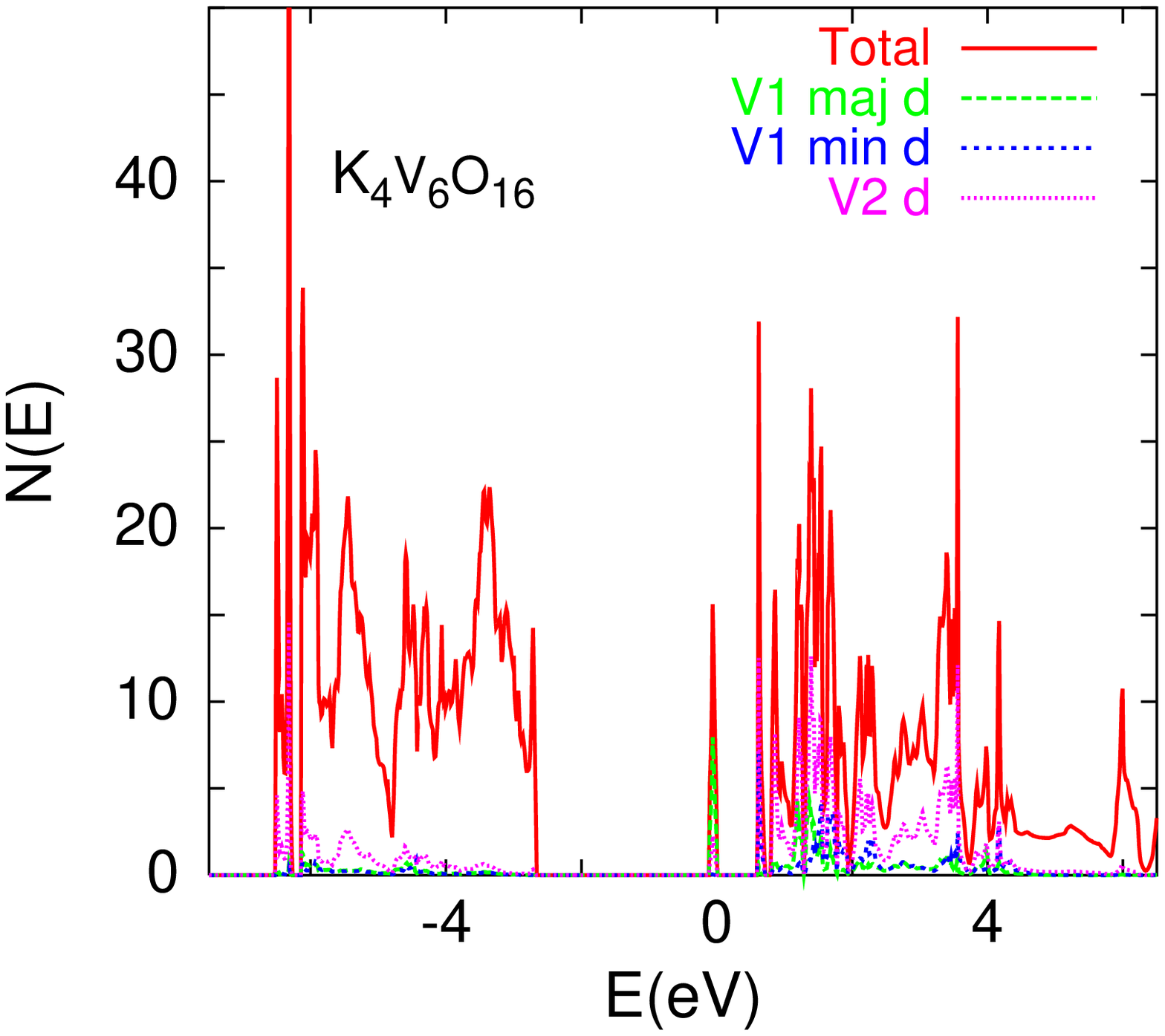,width=0.80\linewidth,angle=0,clip=}}
\centerline{
\epsfig{file=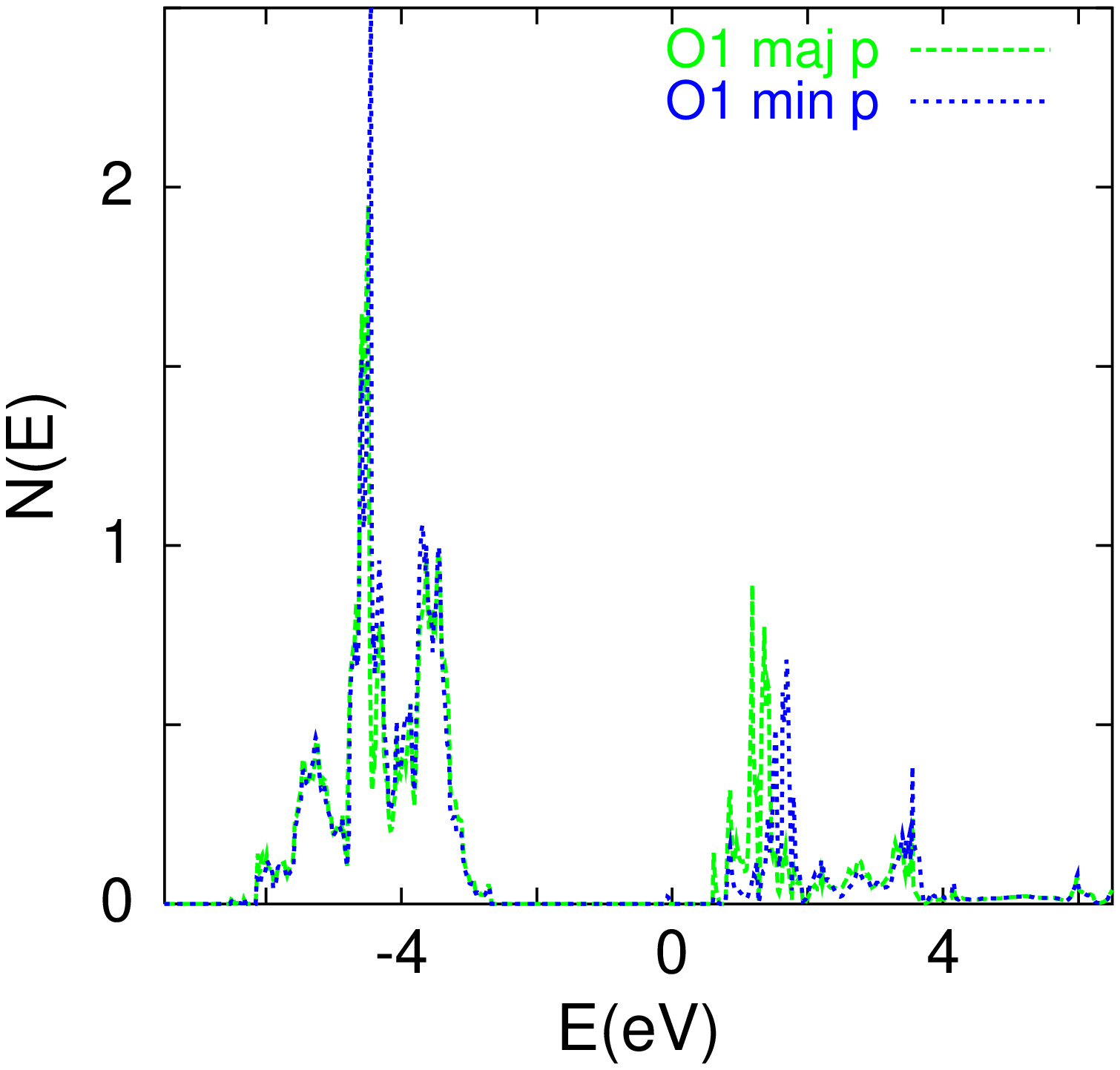,width=0.82\linewidth,angle=0,clip=}}
\vspace{0.2cm} \caption{(Color online) Projections of the electronic
density of states onto the various atomic sites, as measured by
projections onto the corresponding LAPW spheres. The sphere radii
are 1.68 Bohr and 1.35 Bohr for V (top) and O (bottom),
respectively.} \label{dos}
\end{figure}


 The band structure for the antiferromagnetic ground state is shown
in Fig. \ref{bands}. Projections of the density of states onto the V
and the O1 (apical O above V1) LAPW spheres are shown in Fig.
\ref{dos}. The electronic structure shows a main gap of 3.5 eV,
which separates a manifold of occupied O 2$p$  derived bands from an
unoccupied conduction band manifold, which derived from V 3$d$
states near the bottom. There are two bands in the gap. These are
the exchange split vanadyl majority and minority spin states. These
are very flat bands, and show practically the same dispersion,
reflecting the weak interaction of these split off states with the
other orbitals. The LSDA exchange splitting of the vanadyl magnetic
$d$ state is 0.73 eV. The band gap is smaller (0.58 eV) due to the
$\sim$ 0.15 eV width of these bands. However, optical transitions at
this energy would be forbidden both because of the indirect nature
of the gap, and more importantly because this is a spin-flip
transition. The lowest allowed transitions would be from the
majority spin vanadyl state to the edge of the main conduction
manifold, starting at 0.90 eV. Since the unoccupied, minority spin
vanadyl state lies 3.4 eV above the main O $2p$ band edge, all
optical transitions up to 3.4 eV are from the polarized majority
vanadyl 3$d$ state to the conduction bands, and are in the majority
spin channel only. As may be seen from the projections of the
density of states, the occupied vanadyl state (this is the $d$ state
of the crystal field 3$d$ shell of V in the presence of the very
short V1-O1 bond), has practically pure V1 $d$ character, as
expected. However, the higher lying V$d$ states have substantial
hybridization with O $2p$ orbitals. Transitions from the main O $2p$
derived valence band manifold to the main conduction bands, which
would be expected to yield a strong optical signal, would begin at
3.5 eV.

 This picture is reflected in the calculated LSDA optical
conductivity, shown in the lower inset of Fig. \ref{Conductivity}. A
broad feature is found extending from $\sim$0.9 to 2.0 eV, derived
from two strong peaks at 0.9 eV and 1.5 eV, plus three smaller
peaks, with a strong edge beginning at $\sim 3.5$ eV, and a
prominent peaks at 4.0 and 4.6 eV. While the agreement with
experiment is by no means perfect, the comparison does allow
identification of the main features. In particular, the features can
be mapped onto the experiment by noting that the main peaks are
shifted consistent with a downward shift of the majority occupied
$d$ level in the experiment by approximately 0.5 eV, relative to the
LSDA band structure, perhaps due to Mott-Hubbard correlations.
Allowing for such a shift, the LSDA optical conductivity is more
similar to the 300 K data than the 4 K data, which presumably
reflects the effect of the structural transition at 110 K. In any
case, the comparison allows us to identify the magnetochromic
features in the optical spectrum as deriving from transitions in the
majority spin only between the occupied V1 $d$ state and hybridized
V-O conduction band states. One test of this basic picture of the
electronic structure would be to measure the transport gap, which
should be smaller than the optical gap in the scenario above.

\subsection{Magneto-Dielectric Properties of K$_2$V$_3$O$_8$}

\begin{figure}[ht]
\includegraphics[width=3.0in]{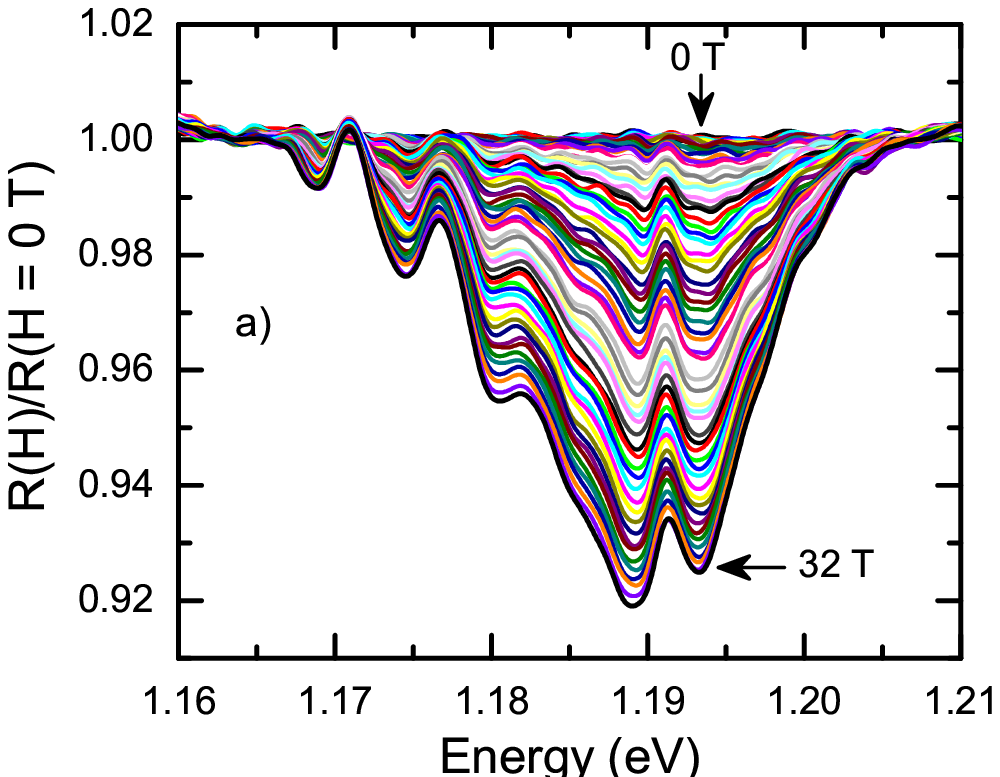}
\includegraphics[width=3.0in]{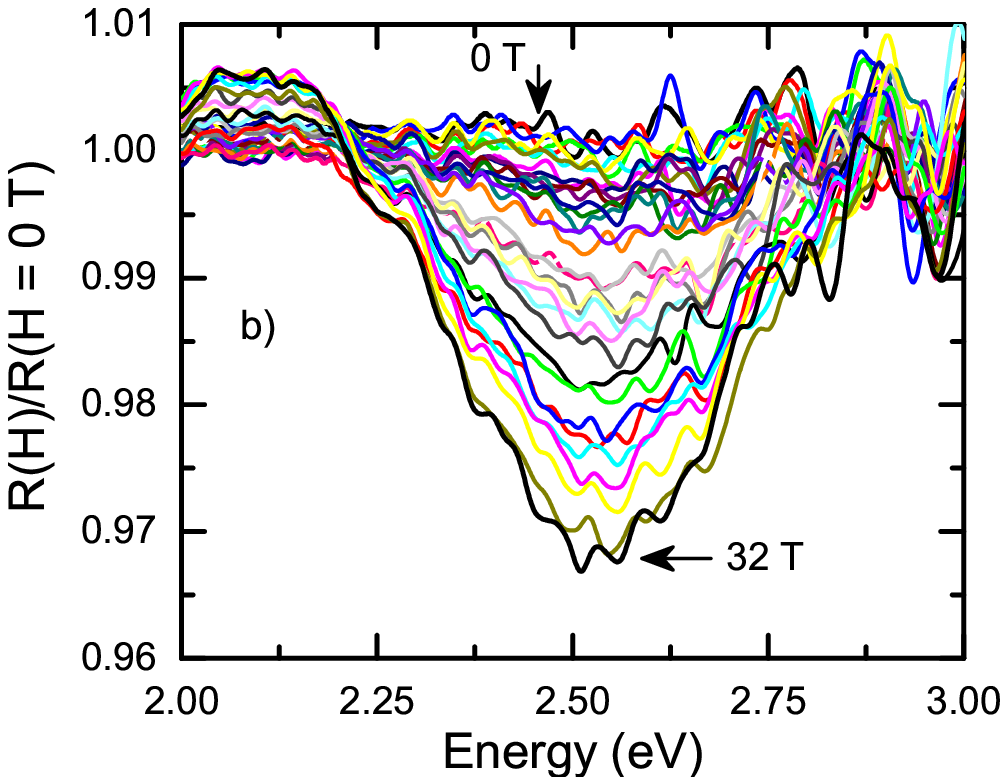}
\vspace{-0.4cm} \caption{(Color online) The normalized $ab$-plane
magneto-optical response, R($H$)/R(0 T), of K$_2$V$_3$O$_8$ for
magnetic fields between  0 and 32 T ($H$$\parallel$$c$) at 4.2 K (a)
A sharp structure centered at 1.19 eV is observed in the near
infrared. Data are taken with 0.5 T steps to highlight the
multiphonon coupling. (b) A broad structure centered at 2.5 eV is
observed in the color band region. Data are taken with 1 T
steps.}\label{magnetoboth}
\end{figure}

Figure \ref{magnetoboth} displays a close-up view of the
magneto-optical response of K$_2$V$_3$O$_8$, R($H$)/R(0 T) for
$H$$\parallel$$c$, at 4.2 K.\cite{note1} The applied magnetic field
decreases the overall reflectance in a significant and systematic
way in both the near infrared (at $\sim$1.19 eV) and in the optical
regime (centered at $\sim$2.5 eV). The feature centered at 1.19 eV
(Fig. \ref{magnetoboth}(a)) is sharp and displays a great deal of
fine structure that increases in amplitude with applied field,
whereas the feature near 2.5 eV (Fig. \ref{magnetoboth}(b)) is very
broad. Based upon the energy scale of these magneto-optical effects,
both structures are associated with field-induced changes in the
$V^{4+}$ $d$ $\rightarrow$ $d$ on-site excitations. We refer to this
type of effect as ``magnetochromism" when it occurs in the visible
range.

\begin{figure}[t!]
\includegraphics[width=3.0in]{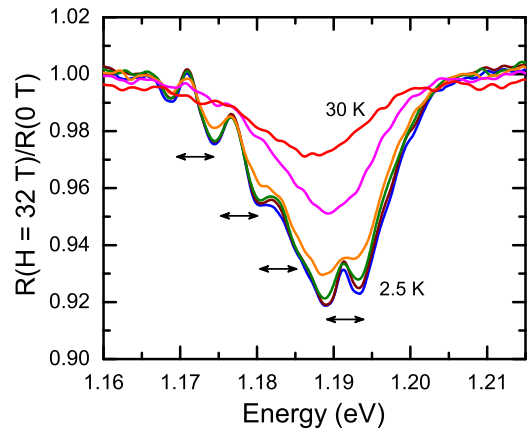}
\vskip -0.2cm \caption{(Color online) Magneto-optical response of
K$_2$V$_3$O$_8$ in the $ab$-plane for $H$ = 32 T at 2.5, 4.2, 7, 10,
20, and 30 K (bottom to top). Arrows indicate the fine structures
with energy scale $\sim$6.8 meV (55 cm$^{-1}$).}\label{TempDep32T}
\end{figure}

Figure \ref{TempDep32T} shows the detailed temperature dependence of
the 1.19 eV reflectance ratio feature at $H$ = 32 T. Here, the
overall effect increases with decreasing temperature, saturating
near 4 K ($\approx$$T_N$). The 1.19 eV reflectance ratio structure
is punctuated by several sidebands, each spaced by $\sim$6.8 meV (55
cm$^{-1}$). This is a vibrational energy scale. Thus, these
sidebands provide  experimental evidence that lattice coupling may
be important in K$_2$V$_3$O$_8$. Interestingly, the prominent low
temperature vibrational fine structure on the 1.19 eV reflectance
ratio feature disappears through the 15 K transition, which is
associated with formation of magnetic correlations in this highly 2D
system. Degradation of short-range spin correlations also gives rise
to a peak in the thermal conductivity and magnetic
susceptibility.\cite{Lumsden2001}

\begin{figure}[b!]
\includegraphics[width=3.0in]{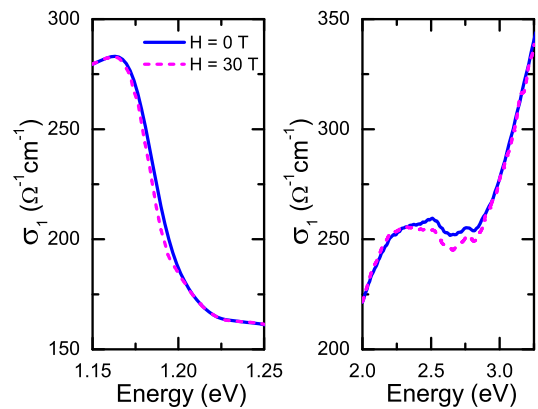}
\vskip -0.2cm \caption{(Color online) The $ab$-plane optical
conductivity spectra of K$_2$V$_3$O$_8$  at $H$ = 0 T (solid line)
and $H$ = 30 T (dashed line) at 4.2 K.}\label{Fieldconds}
\end{figure}

How do these trends manifest themselves in the optical properties?
Since $\epsilon(\omega)=\epsilon_1(\omega) + i\epsilon_2(\omega)
=\epsilon_1(\omega) + {{4{\pi}i}\over {\omega}}\sigma_1(\omega)$, it
is clear that the field-induced changes in reflectance, discussed in
previous paragraphs, translate into finite frequency
magneto-dielectric effects. For instance, to obtain the 30 T optical
conductivity, we renormalize the zero-field absolute reflectance
with the high-field reflectance ratio, and calculate $\sigma$$_1$
using Kramers-Kronig techniques.\cite{wooten1972} The results are
shown in Fig. \ref{Fieldconds}.

The panels in Fig. \ref{Fieldconds} display close-up views of the
optical conductivity of K$_2$V$_3$O$_8$ at 0 and 30 T. Based upon
these results, we can directly attribute the observed
magneto-optical response to field-induced modifications of the
$V^{4+}$ $d$ $\rightarrow$ $d$ excitations. The main effect of the
applied field is to red-shift the trailing edge of the 1 eV
excitation and to narrow the color band excitation centered at 2.3
eV. Analysis of the partial sum rule (not shown) indicates that
these excitations narrow in magnetic field, perhaps due to a change
in scattering rate; a very small amount of oscillator strength is
missing at 30 T. One plausible origin for the modification of the
$V^{4+}$ $d$ $\rightarrow$ $d$ on-site excitations is a
field-induced structural deformation of the VO$_5$ square pyramid.
In this system, displacement of the vanadium ion in the square
pyramids (toward the apical oxygen and out of plane) gives a very
short V-O (vanadyl-type) bond. Given the evidence for a soft lattice
(detailed below), distortion of the local crystal field environment
in the VO$_5$ square pyramid may be driven by a magneto-elastic
coupling mechanism. The fact that the optical transitions below 3.5
eV are due to transitions from a nearly pure spin polarized V d
state to hybridized states, involving V and O states, also suggests
the possibility of strong lattice coupling.

\begin{figure}[t]
\includegraphics[width=3.0in]{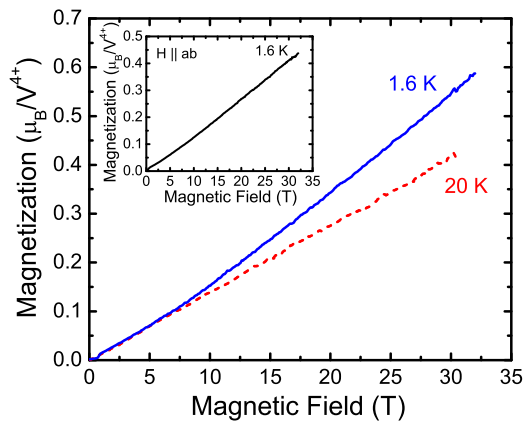}
\vskip -0.2cm \caption{(Color online) Magnetization of
K$_2$V$_3$O$_8$ as a function of applied magnetic field
($H$$\parallel$$c$) at 1.6 K (solid line) and 20 K (dashed line).
The inset shows the magnetization for $H$$\parallel$$ab$ at 1.6 K.
On this scale, magnetization will saturate at 1
${\mu_B/V^{4+}}$.}\label{magnetization}
\end{figure}

To complement our magneto-optics results, we investigated the
high-field magnetization of K$_2$V$_3$O$_8$ (Fig.
\ref{magnetization}). At 20 K, the magnetization increases more or
less linearly up to at least 33 T due to the gradual alignment of
spins towards the applied field direction by tilting out of the
basal plane.\cite{spinflop} A careful inspection of the low
temperature data reveals nonlinear behavior for both
$H$$\parallel$$c$ and $H$$\parallel$$ab$. As shown in Fig.
\ref{magnetization}, the slope $dM/dH$ clearly increases above 9 T
at 1.6 K, with no sign of saturation up to 33 T for both $H$$\|$$c$
and $H$$\|$$ab$. Apparently, the high magnetic field reduces the
antiferromagnetic coupling strength at low temperatures. Based on
the work of Chernyshev,\cite{Chernyshev05} we estimate the canting
angle to be $\sim$29$^\circ$  at 33 T ($H$$\|$$c$). Besides the
change in slope at 9 T, there are no other features in the
magnetization that correlate with the optical properties.

\begin{figure}[h!]
\includegraphics[width=3.3in]{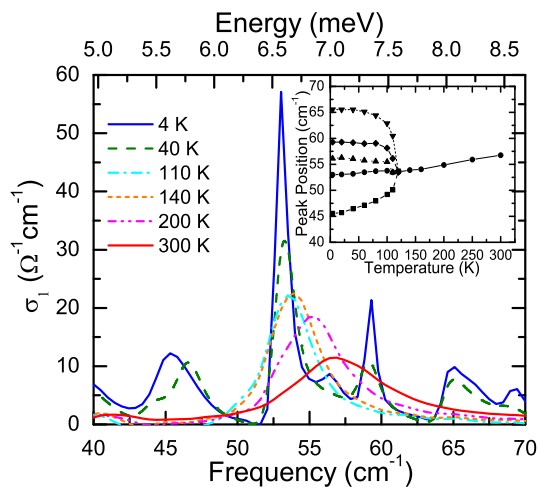}
\vskip -0.2cm \caption{(Color online) Peak splitting in the far
infrared spectra of K$_2$V$_3$O$_8$ at different temperatures. The
inset shows the center frequency of the 55 cm$^{-1}$ vibrational
mode versus temperature.}\label{Phonons55}
\end{figure}

 In order to further elucidate the role of the lattice in
 mediating the magneto-dielectric response, we measured the variable
temperature vibrational properties of K$_2$V$_3$O$_8$. Several modes
show softening and splitting at low temperature, evidence of the 110
K structural distortion and a soft lattice. Based on vibrational
properties studies in similar vanadyl oxide
compounds,\cite{Popov2000,Popov2002,Konst2000,Konst2002,Spitaler2004}
the features of interest below 500 cm$^{-1}$ can be assigned as
V-O-V bending, O-V-O bending, and other low-energy tortional and
twisting motions of the square pyramidal and tetrahedral building
block units. The behavior of the 6.8 meV (55 cm$^{-1}$) cluster
(Fig. \ref{Phonons55})  is typical of the low temperature mode
splitting in K$_2$V$_3$O$_8$. Similar splitting is observed near
17.4 meV (140 cm$^{-1}$) (doublet), 45.3 meV (365 cm$^{-1}$)
(quintuplet), and 52.9 meV (427 cm$^{-1}$) (doublet). Interestingly,
the energy scale of the 6.8 meV (55 cm$^{-1}$) mode matches that of
the multiphonon fine structure observed at 1.19 eV in the
magneto-optical response (Fig. \ref{magnetoboth}(a)). To check for
direct evidence of magneto-elastic interactions, we carried out
independent magneto-infrared measurements of K$_2$V$_3$O$_8$. There
is no magnetic field dependence of the 6.8 meV mode up to 17 T,
within our sensitivity. That said, it is important to note that
K$_2$V$_3$O$_8$ has many infrared, Raman, and inactive vibrational
modes. Raman-active torsional modes that flex the VO$_5$ square
pyramids out of the plane at this energy scale would seem to be
likely candidates for coupling to the magnetic system. We carried
out LSDA calculations of some of the zone center phonons, and find
modes of this character in the appropriate frequency range.
SrCu$_2$(BO$_3$)$_2$ also has similar Raman-active out-of-plane
motions of almost all ions that couple to the magnetic
system.\cite{KChoi2003} The coupling of a soft phonon mode (or other
lattice degrees of freedom) with localized spins is responsible for
magneto-dielectric effects in  antiferromagnets such as EuTiO$_3$,
DyMn$_2$O$_5$, and HoMnO$_3$ as
well.\cite{Katsu2001,NHur2004,Cruz2005} Future work will focus on
the connection between high and low frequency magneto-dielectric
effects in K$_2$V$_3$O$_8$ and other materials.

\section{Conclusion}

We investigated the optical, magneto-optical, and LSDA electronic
structure properties of K$_2$V$_3$O$_8$, a model $S$=1/2
quasi-two-dimensional Heisenberg antiferromaget. The spectral
response is similar to that of other inhomogeneously mixed-valent
vanadates, with a 0.5 eV semiconducting gap and both $V^{4+}$ $d$
$\rightarrow$ $d$ on-site and O 2$p$ $\rightarrow$ V 3$d$ charge
transfer excitations. According to our electronic structure
calculations, the on-site $d$ $\rightarrow$ $d$ transitions emanate
from strongly spin polarized (majority channel) excitations.
Application of magnetic field modifies the optical properties in two
regimes: (1) on the trailing edge of the 1 eV excitation and (2) in
the shape of the color band near 2.5 eV. These effects are
 attributed to changes in the $V^{4+}$ $d$ $\rightarrow$ $d$
on-site excitations due to field-induced local distortion of the
VO$_5$ square pyramid. Combined with evidence for a soft lattice,
the presence of vibrational fine structure on the sharp 1.19 eV
magneto-optical feature, and the fact that these optical excitations
are due to transitions from a nearly pure spin polarized V $d$ state
to hybridized states involving both  V and O suggest that the
magneto-dielectric effect in K$_2$V$_3$O$_8$ may be driven by strong
lattice coupling. We discussed the type of low-energy vibrational
modes that might couple the VO$_5$ square pyramids to the magnetic
system.

\section{ACKNOWLEDGMENTS}

Work at the University of Tennessee is supported by the Materials
Science Division, Basic Energy Sciences, U.S. Department of Energy
(DE-FG02-01ER45885). Oak Ridge National Laboratory is managed by
UT-Battelle, LLC, for the U.S. Dept. of Energy under contract
DE-AC05-00OR22725. A portion of this work was performed at the
NHMFL, which is supported by NSF Cooperation Agreement DMR-0084173
and by the State of Florida. We thank Igor I. Mazin for interesting
discussions and Sonal Brown for technical assistance.

\vfill\eject


\begin{thebibliography}{RamRai}
\bibitem{synthesis}
About 100 grams of V$_2$O$_5$ (99.995 $\%$) and 76 grams of
K$_2$CO$_3$ (99.997$\%$) is loaded into a 250 ml Pt crucible and
slowly heated in air to 700 $^\circ$C and held for 2 hours. The
powder is added in two stages because of substantial foaming. The
molten KVO$_3$ flux is cooled to room temperature and the Pt
crucible and solidified KVO$_3$ is loaded into a fused silica
ampoule that is necked down at the top. Using a long funnel, 9 grams
of VO$_2$ (99 $\%$) is added to the Pt crucible and the entire
ampoule is evacuated and sealed. The sealed ampoule is loaded into a
furnace and heated to 850 $^\circ$C for 6 hours, cooled to 700
$^\circ$C over 1h and then cooled to about 400 $^\circ$C at 1
$^\circ$C/h, followed by furnace cooling to room temperature. The
KVO$_3$ flux is removed with a combination of warm water and
ultrasonic vibration. The resulting K$_2$V$_3$O$_8$ crystals are
black rectangular plates with typical dimensions of 5 x 5 x 1
mm$^3$.
\bibitem{code-note}
Calculations were done with two different LAPW codes, an in-house
LAPW code, and the WIEN2K code. These were cross-checked and found
to yield the same properties.
\bibitem{note1}
The magneto-optical effects discussed here are reproducible on the
same crystal and crystals of different batches. There is no
hysteresis in the magneto-optical response, which implies that the
field-induced structural deformations are reversible.
\end{thebibliography}
\end{document}